\documentstyle[11pt,a4wide]{article}
\newcommand{\nl}{ {\hfill \break} }
\newcommand{\np}{ {\newpage } }

\newcommand{\Iff}{ {\Leftrightarrow } }

\newcommand{\imp}{ {\Rightarrow } }

\newcommand{\eps}{ \mbox{$\varepsilon$} }
\newcommand{\cl   }{ \mbox{${\rm cl   }$} }

\newcommand{\trace}{ \mbox{${\rm tr   }$} }

\newcommand{\ad}{ \mbox{\rm ad} }
\newcommand{\spn}{ \mbox{\rm span} }
\newcommand{\N    }{ \mbox{${\rm I\!N    }$} }
\newcommand{\R    }{ \mbox{${\rm I\!R    }$} }
\def\C{\mbox{\rm {I\kern-.520em C}}}
\newcommand{\unity}{ \mbox{{\rm $1$\hspace*{-0.1cm}I    }} }
\newcommand{\GL   }{ \mbox{${\rm GL   }$} }

\newcommand{\End}{ \mbox{${\rm End}$} }
\newcommand{\Aut}{ \mbox{${\rm Aut}$} }
\newcommand{\e}{ \mbox{${\rm e}$} }
\newcommand{\su}{ \mbox{${\rm su}$} }
\newcommand{\so}{ \mbox{${\rm so}$} }
\newcommand{\iso}{ \mbox{${\rm iso}$} }
\newcommand{\cyclic}{ \mbox{{\tiny\rm cyclic}} }
\newcommand{\im}{ \mbox{${\rm im}$} }
\newcommand{\I   }{ \mbox{${\rm I   }$} }
\newcommand{\II  }{ \mbox{${\rm II  }$} }
\newcommand{\III }{ \mbox{${\rm III }$} }
\newcommand{\IV  }{ \mbox{${\rm IV  }$} }

\newcommand{\V   }{ \mbox{${\rm V   }$} }
\newcommand{\VI  }{ \mbox{${\rm VI}  $} }
\newcommand{\VIo }{ \mbox{${\rm VI}_0$} }

\newcommand{\VII }{ \mbox{${\rm VII} $} }
\newcommand{\VIIo}{ \mbox{${\rm VII}_0$} }

\newcommand{\VIII}{ \mbox{${\rm VIII}$} }
\newcommand{\IX  }{ \mbox{${\rm IX  }$} }

\renewcommand{\theequation}{\thesection.\arabic{equation}}

\newcommand{\beq}[1]{\begin{equation}\label{#1}}
\newcommand{\eeq}{\end{equation}}
\newcommand{\bear}[1]{\begin{eqnarray}\label{#1}}
\newcommand{\ear}{\end{eqnarray}}
\newcommand{\nn}{\nonumber}

\begin{document}
\begin{tabbing}
\` {\small Preprint IPM-95}
\end{tabbing}
\vspace{0.7truecm}
\begin{center}
{\Large{\bf
{Classifying spaces for homogeneous manifolds} 
\vspace{0.2cm}\\ 
{and their related Lie isometry deformations\dag}
\vspace{0.2cm}
}}
\end{center}                         
\begin{center}
{\bf {Martin Rainer}}
\ddag 
\vspace{0.5truecm}\\
Gravitationsprojekt / Projektgruppe Kosmologie\\ 
Institut f\"ur Mathematik \\ 
{Universit\"at Potsdam, Am Neuen Palais 10}\\  
{PF 601553, D-14415 Potsdam, Germany}\\
{e-mail: mrainer@aip.de}
\vskip 0.5cm
\end{center}
\begin{abstract}

Among plenty of applications, low-dimensional homogeneous spaces appear
in cosmological models as both, 
classical factor spaces of multidimensional geometry and 
minisuperspaces in canonical quantization. 


Here a new tool to restrict their continuous deformations is presented:
Classifying spaces for homogeneous manifolds 
and their related Lie isometry deformations.


The adjoint representation of $n$-dimensional real Lie algebras induces a 
natural topology $\kappa^n$ on their classifying space $K^n$. 
$\kappa^n$ encodes the natural algebraic relationship between different 
Lie algebras in $K^n$. For $n\geq 2$ this topology is not Hausdorffian. 
Even more it satisfies only the separation axiom $T_0$, but not $T_1$,
i.e. there is a constant sequence in $K^n$ which has a limit different from 
the members of the sequence. Such a limit is called a transition.


Recently it was found that transitions in $K^n$ are the natural generalization 
and transitive completion of the well-known In\"on\"u-Wigner contractions. 
For $n\leq 4$ the relational classifying spaces $(K^n,\kappa^n)$ are 
constructed explicitly.


A Lie algebra $A_n$ of low dimension $n\leq 4$ naturally acts as an isometry 
$G_n=e^{A_n}$ of some homogenous space; so (locally) a homogeneous 
Riemannian $3$-space is either of Kantowski-Sachs type, 
with a transitive $G_4$, or it corresponds to one of the Bianchi types with
a transitive $G_3$.                                                                                                                                                                               


Calculating their characteristic scalar invariants
via triad representations of the characteristic isometry,
local homogeneous Riemannian $3$-spaces are 
classified in their natural geometrical relations to each other. 
Their classifying space is a composition of pieces with 
different isometry types. 
Although it is Hausdorffian,
different $\kappa^3$-transitions to the same limit in $K^3$ 
may induce locally non-Euclidean regions (e.g. at Bianchi $\rm VII_0$).
\end{abstract}
\vspace{0.7truecm}
\dag
{\footnotesize
Lecture given at the $7^{th}$ Regional Conference on Mathematical Physics
\nl
"Caspian Conference", October 15-22, 1995 
}
\vspace{.17truecm}
\nl
{\small\em  
This lecture was financially supported by the DAAD.
}
\vspace{.17truecm}
\nl
{\ddag 
\footnotesize
Present address: 
{\bf I}nstitute for Studies in {\bf P}hysics and {\bf M}athematics\\
Farmanieh Bld., Tehran P.O.Box 19395-5746, Tehran, Iran \\
e-mail: mrainer@physics.ipm.ac.ir 
}
\np
\section{\bf Introduction}  
\setcounter{equation}{0}
Classical cosmological models are formulated in terms of homogeneous
Riemannian $3$-manifolds evolving through a $1+3$-dimensional space-time.
They can be naturally generalized to hypersurface homogeneous models
of arbitrary dimension, including additional internal spaces, each of which 
is homogeneous and Riemannian. A multidimensional model is one
for which all higher dimensional hypersurfaces decompose into
a direct product of Riemannian factor spaces, with the same
fixed dimensions on all hypersurfaces. (A generalization with
dynamical dimensions has been considered recently in \cite{BlMR}.)
For a hypersurface homogeneous multidimensional model, the scales
of the homogeneous factor spaces
provide the only dynamical degrees of freedom. 
They coordinize the minisuperspace, which in this case is a 
Minkowski space with 
dimension equal to the number of Riemannian factor spaces of the
underlying multidimensional hypersurface.
In a more general model,
the homogeneous factor spaces might no longer be independent, 
but their scales may interact locally. Then, the minisuperspace
of this generalized multidimensional model is a homogeneous
Lorentzian manifold. 

Hence, a classification of local homogeneous manifolds of given
dimension, is a clue to a systematic understanding of the possibilites
for dynamical deformations of both, the factor space geometries
of homogeneous generalized multidimensional cosmological models
and the minisuperspace in canonical quantization.
In the former application, the relevant manifolds are Riemannian,
in the latter one Lorentzian. In general however, the problem to
determine the proper classifying spaces is too hard to be solved. 
Therefore in this work, we will restrict to
a complete description just for local Riemannian $3$-manifolds.
(In \cite{MoRSch}
the Lorentzian $3$-manifolds were examined analogously,
for specific orientations of their null hypersurfaces.)

Since the local isometry subgroups of a local homogeneous 
manifold are Lie groups, it is useful, before trying to classify the 
homogeneous manifolds, to find first all possible
contractions and, more generally, all possible limit transitions 
between real (or complex) {Lie} algebras of fixed
dimension, and to uncover 
the natural topological structure
of the space of all such {Lie} algebras.
The topology is given by the algebraic properties of the  {Lie} algebras.
The space $W^n$ of all structure constants of real 
$n$-dimensional {Lie} algebras 
carries the subspace topology induced from the Euclidean $\R^{n^3}$  
(see \cite{Se}). 
The quotient topology $\kappa^n$, obtained from this topology w.r.t. 
equivalence by $\GL(n)$ isomorphisms, renders 
the space $K^n$ of all $n$-dimensional Lie algebras
into a $T_0$ topological space, which is not $T_1$ for $n\geq 2$.
This non-$T_1$ topology has first been described in \cite{Sch}. 
In Sec. 3 below, we derive it with some new method, employing an index
function on the algebra. This approach is somehow inspired by 
Morse theory.
For $n\geq 2$, the space $K^n$ contains some
non-closed point $A$, which has a special limit,
 to another point $B$. 
The inverse of such a transition from $A$ to $B$ is a deformation
of the algebra of $B$ into the algebra $A$.
Note that, unlike in \cite{Rai1,Rai2},
here transitions are defined to include also 
trivial constant limits. 
This has the advantage that also a trivial  
contraction (e.g. in the sense of In\"on\"u-Wigner)
is a transition. This definition is fully compatible with a
partial order  $A\geq B$, which is taken to be the 
specialization order already  used in \cite{Rai1,Rai2}.
This choice of partial order is naturally related to a
Morse like potential $J$, decomposing $K^n$ into subsets   
of different level.

Once the structure of the classifying space $K^n$ is known,
this information can be used as a first ingredient to construct
the space of local Riemannian $n$-manifolds. This is demonstrated
explicitly for $n=3$ below. 
The paper is organized by the following sections.

Sec. 2 reviews the relevant local features of homogenous spaces.
Sec. 3 derives and describes the classifying spaces $(K^n,\kappa^n)$  
of $n$-dimensional Lie algebras. 
An index function $J: K^n\to\N_0$ is related naturally
to the topology $\kappa^n$. 

Sec. 4 shows how the well known In\"on\"u-Wigner \cite{InW} and Saletan
\cite{Sal} contractions correspond to special transitions.
Unlike general transitions, the In\"on\"u-Wigner contractions
are well studied in the literature (e.g. up to real dimension $4$
they have been classified in \cite{Hud}).

In Sec. 5 we review the Bianchi Lie algebras, which constitute the
elements of $K^3$. 
Using the index $J$ of Sec. 3, the topology $\kappa^3$ is described  
explicitly for both, the real and complex case. 

Sec. 6 relates the topology $\kappa^n$ to the Zariski topology,
and explains, via Lie algebra cohomology, why semisimple
Lie algebras, and more generally all rigid ones, do not
admit deformations in the category given by $K^n$.
  
Sec. 7 then classifies all local homogeneous Riemannian $3$-manifolds 
using their isometries in $K^3$,
and Sec. 8 discusses the results in the light of possible consequences
for an evolution of the characteristic symmetries of fundamental 
theories like quantum gravity and cosmology.
\section{\bf Homogenous manifolds}  
\setcounter{equation}{0}
Here we want to investigate the data which characterizes 
the {\em local} structure of a homogeneous Riemannian or pseudo Riemannian
manifold $(M,g)$. If we consider  for arbitrary dimension $n$ 
the different possible signatures modulo the reflection
$g\to-g$, then $1+[\frac{n}{2}]$ different signature classes 
are distinguished by the codimension $s=0,\ldots,n-[\frac{n}{2}]$ 
of the characteristic null hypersurface in the tangent space.
For Lorentzian signature $s=1$ the latter is an $(n-1)$-dimensional
open double cone at the basepoint, 
in the Riemannian case $s=0$ it is just the basepoint itself.

Per definition, a homogeneous manifold admits a transitive action
of its isometry group. Let us restrict here to the case
where it has even more a simply transitive subgroup of the isometry
group. In this case we can solder the metric to an orthogonal frame 
spanned by the Lie algebra generators $e_i$ in the tangent space, i.e.
\bear{solder}
g_{\mu\nu}=e^a_\mu e^b_\nu g_{ab}
\ear
where $e^a=e^a_\mu dx^\mu=g^{ai}e_i$, 
$e_i=e^\mu_i \frac{\partial}{\partial x^\mu}$, 
$g^{ab}g_{ij}=\delta^a_i\delta^b_j$, with the
constant metric
\bear{gab}
\left(g_{ab}\right)=
\left[
\begin {array}{ccc} 
\epsilon_1e^{s}&0&0\\\noalign{\medskip}\
0&{\epsilon_2 e^{s+w-t}}&0\\\noalign{\medskip}
0&0&{\epsilon_3e^{s-t}}
\end{array}
\right].
\ear
Here $s$ fixes the overall scale, while $t$ and $w$ parametrize
the anisotropies related respectively to the $e_1$ and $e_2$ direction
(maintaining isotropy in the respective orthogonal planes).   

The local data can be rendered in form of (i) the 
{\em local scales} of (\ref{gab}),
(ii) the {\em covariant derivatives}
\bear{De}
D e^k=e^k_{i;j} e^i e^j:= e^k_{\alpha;\beta} dx^\alpha dx^\beta
\ear
of the dual generators $e^k$ in the cotangent frame, 
(iii) the corresponding {\em Lie algebra}
\bear{Lie}
[e_i,e_j]=C^k_{ij} e_k,
\ear
and (iv) the {\em orientation} 
of the $n-s$-dimensional null hyperspace in the 
tangent space. 
For the Lorentzian case $s=1$, this orientation is described
by the future oriented normal vector $n$ along the central axis
of the double cone, 
\bear{or}
n=n^a e_a,
\ear
where its triad frame components $n^a$ have to be coordinate independent,
since the manifold is assumed to be homogeneous.  

Let us consider now $n=3$. In this case, 
there are no further signature cases besides the 
Riemannian and Lorentzian ones. 
For the $3$ special cases $n^a=\delta^a_i$, $i=1,\ldots,3$
an explicit description of the Lorentzian $3$-spaces of nonflat
Bianchi type has been given in \cite{MoRSch}.
However a complete classification of all homogeneous Lorentzian $3$-spaces                                                                                                                  
need to control systematically the effect of different orientations
(\ref{or}). Presently, this problem still remains to be solved.

Therefore let us consider in the following only the Riemannian case,
where the datum (iv) is trivial. Sec. 8 will give the complete 
classification of local homogeneous $3$-spaces with
some isometry subgroup in $K^3$.
Moreover, the Kantowski-Sachs (KS) spaces,
here the only exception not admitting a simply-transitive subgroup of their
isometry group, can be obtained as a specific limits of Bianchi IX spaces.
The global geometrical correspondence of such a limit is given 
by a hyper-cigar like $3$-ellipsoid of topology $S^3$, 
stretched infinitely long to become a hyper-cylinder $S^2\times\R$.
So finally we will have a classification  of {\em all}  local
homogeneous Riemannian $3$-manifolds.
\section{\bf Classifying spaces $K^n$ of $n$-dimensional Lie algebras}  
\setcounter{equation}{0}
A (real) {\em (finite-dimensional) Lie algebra}
is a (real) vector space $V$ of dimension $n$, equipped with a 
skew symmetric bilinear product $[\cdot,\cdot]$, 
satisfying the {\em Jacobi condition} $[[x,y],z]+[[y,z],x]+[[z,x],y]=0$ 
for all $x,y,z\in V$.
The evaluation of the Lie bracket  $[\cdot,\cdot]$ on a complete 
set of basis vectors $\{e_i\}_{i=1,\ldots,n}$ 
yields a description of the Lie algebra by a set of 
structure constants $\{C^k_{ij}\}_{i,j,k=1,\ldots,n}$ from
Eq. (\ref{Lie}). Equivalently the
endomorphisms $C_{i}:=\ad(e_i)$, $i=1,\ldots,n$, 
{}from the adjoint representation
$\ad: e_i\to [e_i,\cdot]$, carry the same information
on the algebra.
Note that this description is overcomplete:
Due to its antisymmetry, the Lie algebra is already
completely described by the $(n-1)\times (n-1)$-matrices
$C_{<i>}$, $i=1,\ldots,n$, each with components 
$C^k_{ij}$ ${j,k=1,\ldots,n-1}$.  
But, as we will see, also this description may still carry redundancies. 

The bracket $[\cdot,\cdot]$ defines a {Lie} algebra, 
iff the structure constants
satisfy the $n\{{n\choose 2}+{n\choose 1}\}$ antisymmetry 
conditions
\begin{equation}\label{anti}
C^k_{[ij]}=0,
\end{equation}
and, corresponding to the Jacobi condition, 
the $n\cdot{n\choose 3}$ quadratic compatibility constraints 
\begin{equation}\label{Jacobi}
C^l_{[ij}C^m_{k]l}=0
\end{equation}
with nondegenerate antisymmetric indices $i,j,k$.

Here we only deal with finite-dimensional Lie algebras.
Hence the adjoint representation in $\End(V)$ 
gives a natural associative matrix representation
of the algebra, generated by the matrices $C_i$.
Using this representation, the associativity of the matrix product
$C_i\cdot C_j$ implies with $[C_i,C_j]=C^k_{ij}C_k$ that 
the Jacobi condition (\ref{Jacobi}) is an {\em identity}
following already from Eq. (\ref{anti}). 
However, if we do not use this extra knowledge from the adjoint 
representation, then, for $n>2$, Eq. (\ref{Jacobi})  yields algebraic
relations independent of Eq. (\ref{anti}).

For $n\geq 2$ there exists an irreducible tensor decomposition $C=D+V$, 
i.e.     
\begin{equation}\label{dec} 
C^k_{ij}=D^k_{ij}+V^k_{ij},
\end{equation}
where $D$ is the tracefree part,  i.e. $\trace(D_i):=D^k_{ik}=0$,  
and $V$ is the vector part, 
\beq{vector} 
V^k_{ij}:=\delta^k_{[i}v_{j]},
\eeq 
given by $v_i:=\frac{2}{1-n}\trace(C_{i})$, $i=1,\ldots,n$. 
The Lie algebra is {\em tracefree} 
(corresponding Lie groups are {\em unimodular}) iff $V\equiv 0$, and it is 
said to be of {\em pure vector type} iff $D\equiv 0$. 
For each $n$, there exists exactly
one non-{Abel}ian pure vector type {Lie} algebra, denoted by $\V^{n}$.
For $n=3$, the latter is the Bianchi type \V,
and the decomposition (\ref{dec}) is given by
\begin{equation}
C^k_{ij}=\eps_{ijl}(n^{lk}+\eps^{lkm}a_m),\quad
D^k_{ij}=\eps_{ijl} n^{lk},\quad
v_i=2 a_i,
\end{equation}
where $n^{ij}$ is symmetric and $\eps^{ijk}$ is the usual antisymmetric 
tensor (cf. also \cite{LL}).
Hence for $n=3$, the Jacobi condition (\ref{Jacobi})
can be written as
\beq{jac3}
n^{lm}a_m=0.
\eeq
These $3$ nontrivial relations are in general independent of 
Eq. (\ref{anti}).

For arbitrary $n$, 
the space of all sets $\{C^k_{ij}\}$ satisfying the {Lie} algebra 
conditions (\ref{anti}) and (\ref{Jacobi}) 
is a subvariety  $W^n \subset \R^{n^3}$, with a dimension
\beq{dimW}
\dim W^n \leq n^3 - \frac{n^2(n+1)}{2}  =\frac{n^2(n-1)}{2},
\eeq
bounded by Eq. (\ref{anti}).
For $n\geq 3$ the inequality is strict, because
(\ref{Jacobi}) is non trivial in general.
For $n=3$,  the bound (\ref{dimW}) reads
$\dim W^n \leq 9$, and taking into account the 
$3$ additional relations of Eq. (\ref{jac3})
actually yields  $\dim W^n = 6$.

$\GL(n)$ basis transformations  act on a given set of structure constants
as $\GL(n)$ tensor transformations:
\begin{equation}
{C}^k_{ij}\to {\tilde C}^k_{ij}:= (A^{-1})^k_h\ C^h_{fg}\ A^f_i\ A^g_j \ \ 
\forall A \in \GL(n).
\end{equation}
On $W^n$ this yields a natural equivalence relation  $C\sim \tilde C$, 
defined by
\begin{equation}\label{equiv}
C^k_{ij} \sim \tilde C^k_{ij} 
:\Iff \exists A \in \GL(n): 
{\tilde C}^k_{ij} = (A^{-1})^k_h\ C^h_{fg}\ A^f_i\ A^g_j, 
\end{equation}
with associated projection $\pi$ to the quotient space, 
\begin{equation}
\pi: \left\{ 
\begin{array}{rcl}
W^n &\to& K^n:=W^n/\GL(n)\\ 
C &\mapsto& [C] 
\end{array}
\right.
\end{equation}
\begin{equation}\label{dimK}
\dim W^n> \dim K^n\geq \dim W^n - n^2.
\end{equation}
The upper bound in Eq. (\ref{dimK}) is a strict one, 
because multiples of $\unity\in\GL(n)$ give 
rise to equivalent points of $W^n$.
Note however that, while, for a given $C\in W^n$, 
certain transformations $A\in\GL(n)$  transform $C\mapsto\tilde C\neq C$,
others keep $C=\tilde C$ invariant. The latter transformations
constitute the automorphism group $\Aut(C)\subset\GL(n)$ of the 
adjoint representation associated with $C$. 
In general, the ${\GL}(n)$ action on $W^n$ is not free,
i.e. there exist points $C$ with $\dim \Aut(C)>0$.
So, Eqs. (\ref{dimW}) and  (\ref{dimK}) provide only very weak
bounds on $\dim K^n$, which is still unknown for general $n$ 
(in the complex case, a more sophisticated upper bound estimate 
has been given in \cite{Ne}). Note also that, e.g. for $n=3$,
the lower bound is trivial, because $\dim W^3 - 3^2=-3<0$.
Actually $\dim \Aut(C)\geq 3$ for all $C\in W^3$.
In general, let us define the {\em automorphic dimension} of $W^n$ as 
\beq{dimAut}
\dim_{\Aut}(W^n):=\min_{C\in W^n}\{ \dim\Aut(C) \}.
\eeq
For any $A\in K^n$, consider a $1$-parameter family of
neighbourhoods $U_{\eps}(A) \subset H(A)$ 
within the Hausdorff connected component $H(A)$
of $A$. Let us define the {\em dimension of the infinitesimal Hausdorff 
connected neighbourhood} of $A$ as 
\bear{dimH}
\dim H(A):=\lim_{\eps\to 0} \dim U_{\eps}(A)
\ear 
Then,  
\bear{estime}
&&\dim W^n=\max_{C\in W^n}\{\dim \pi^{-1}([C]) +\dim H([C]) \}
\nn\\
&&\leq \max_{C\in W^n}\{\dim \pi^{-1}([C])\}+\max_{C\in W^n}\{\dim H([C]) \}
\nn\\
&&=n^2-\min_{C\in W^n}\{ \dim\Aut(C) \} + \dim K^n.
\ear
Using (\ref{dimAut}), the lower bound of Eq. (\ref{dimK}) can be sharpened
yielding
\begin{equation}\label{dimKs}
\dim W^n> \dim K^n\geq \dim W^n - n^2 +\dim_{\Aut}(W^n).
\end{equation}
Note that $\dim\Aut(C)=\dim\Aut([C])$ for any $C\in W^n$.
So $\dim_{\Aut}(W^n)$ actually depends only on $K^n$, and
\beq{dimAutK}
\dim_{\Aut}(W^n)=\min_{C\in W^n}\{ \dim\Aut([C]) \}
=\min_{A\in K^n}\{ \dim\Aut(A) \}=:\dim_{\Aut}(K^n)
\eeq
is the automorphic dimension of $K^n$.

The space $K^n$ of isomorphism classes of $n$-dimensional
Lie algebras is naturally rendered a topological space $(K^n,\kappa^n)$,
where the quotient topology $\kappa^n$ is generated by the projection $\pi$ 
{}from the subspace topology 
on $W^n \subset \R^{n^3}$. 
In order to describe  $(K^n,\kappa^n)$,
let us first recall the axioms of {\em separation}  
(German: {\em Trennung}; cf. e.g. \cite{Ri}):
\nl
{\bf $T_0$}: For each pair of different points there is an open set
containing only one of both.
\hfill\mbox\break 
\nl
{\bf $T_1$}: Each pair of different points has a pair of open 
neighbourhoods with their intersection containing none of both points.
\hfill\mbox\break 
\nl
{\bf $T_2$} ({\em Hausdorff}): 
Each pair of different points has a pair of disjoint neighbourhoods.
\nl 
It holds: $T_2\imp T_1\imp T_0$. 
Often it is more convenient to use the equivalent
characterization of the separation axioms in terms 
of sequences and their limits:
\nl
$T_0$ $\Iff$ For each pair of points there is a sequence converging
only to one of them.
\nl
$T_1$ $\Iff$ Each constant sequence has at most one limit.
\nl
$T_2$ $\Iff$ Each sequence, indexed by a directed partially ordered set, has 
at most one limit.
\nl
$T_1$ is equivalent to the requirement that each $1$-point set is closed.
Actually, for $n\geq 2$, the topology $\kappa^n$ is not $T_1$,
but only $T_0$. This means that there exists some point $A\in K^n$,
which is not closed, 
or in other words, there is a non-trivial {\em transition} from
$A$ to $B\neq A$ in $\cl \{A\}$. 
Non-trivial ($A\neq B$) transitions are special limits, 
which exist  only due to the non-$T_1$ property of $\kappa^n$. 
Here transitions from $A$ to $B$ are defined by
\beq{partial}
A\geq B :\Iff  B\in \cl \{A\}.   
\eeq
By this definition, transitions are transitive 
and yield a natural partial order. A transition
$A\geq B$ is non-trivial, iff $A>B$. 

In the following we want to construct a minimal graph for 
the classifying space $(K^n,\kappa^n)$.
Let us associate an {\em arrow} $A\to B$ to a pair of algebras 
$A,B\in K^n$, with $A>B$, such that
there exists no $C\in K^n$ with $A>C>B$.
We call $A$ the {\em source} and $B$ the {\em target} of
the arrow $A\to B$.  
Now we define a discrete {\em index} function $J: K^n\to\N_0$ as following:
We start with the unique minimal element $\I^n$, to which we assign
the minimal index $J(\I^{n})=0$. Then, for $i\in\N_0$, we assign the 
index $J(S)=i+1$ to the source algebra $S$ of any arrow pointing towards a
target algebra $T$ of index $J(T)=i$, until, eventually 
for some index $J=i_{\max}$ there is no arrow to any target algebra $T$ with 
$J(T)=i_{\max}$. Let us denote
the subsets of all elements with index $i$
as {\em levels} $L(i)\subset K^n$. 

For $n\geq 2$, $K^n$ is directed 
towards its {\em minimal element},
the Abelian Lie algebra $\I^{n}$, constituting its only {\em closed point}. 
For $n\geq 3$ there are points in $K^n$ which are neither open nor closed.

Open points correspond to {\em locally rigid} Lie algebras $C$, 
i.e. those which cannot be deformed to some $A\geq C$ with
index $J(A)>J(C)$. In this sense,
the {\em open points} in $K^n$ are its {\em locally maximal elements}.

Isolated open points 
correspond to {\em rigid} Lie algebras $C$, i.e. those
which cannot be deformed to any $A\in K^n$  with $A\not\leq C$
and index $J(A)\geq J(C)$.
In this sense,
the {\em isolated open points} 
are the {\em locally isolated maximal elements}.
In Sec. 7 below, the isolated open points are considered also from
a dual perspective.

$K^1$ contains only the Abelian algebra $\I^1$. $K^2$ contains
$2$ algebras, the Abelian $\I^2$ and the isolated open
point $\V^2$, with a non-trivial transition $\V^2\to\I^n$.
At the end of Sec. 6 the minimal graph for
$K^3$ is given explicitly. In \cite{Rai1,Rai2} 
the topological structure of $K^4$ (which will not be displayed here)
has been constructed likewise. 

\section{\bf The special transitions of In\"on\"u-Wigner and Saletan}  
\setcounter{equation}{0}
Special kinds of transitions
on a certain 2-point set $\{A,B\}$
of {Lie} algebra isomorphism classes
are the contractions of {In\"on\"u-Wigner} \cite{InW} and their
generalization by {Saletan} \cite{Sal}. 
  
Consider a $1$-parameter set of matrices $A_t\in\GL(n)$ with
$0<t\leq 1$, having a well defined matrix limit 
\beq{Sal}
A_0:=\lim_{t\to 0} A_t
\eeq
which is {\em singular}, i.e. $\det A_0=0$.

For given structure constants $C^k_{ij}$ of a {Lie} algebra
$A$ let us define for 
$0<t\leq 1$ further structure constants 
\beq{eqcontr}
C^k_{ij}(t):=(A^{-1}_t)^k_h\ C^h_{fg}\ (A_t)^f_i\ (A_t)^g_j,
\eeq 
which, according to (\ref{equiv}), all describe the same {Lie} algebra $A$.

If there is a well defined limit $C^k_{ij}(0):=\lim_{t\to 0} C^k_{ij}(t)$,
which satisfies conditions (\ref{anti}) and (\ref{Jacobi}),
yielding well defined structure constants of a {Lie} algebra $B$,
then the associated transition $A\leq B$ 
is called a {\em (Saletan) contraction}.

Moreover a contraction is called 
{\em In\"on\"u-Wigner contraction}
if there is a basis $\{e_i\}$ in which
\beq{IW}
A(t)=
\left(
\begin{array}{cc}
 E_m & 0 \\ 
 0   & t\cdot E_{n-m}              
\end{array}
\right)
\qquad \forall t\in [0,1],  
\eeq 
where $E_k$ denotes the $k$-dimensional unit matrix
(cf. \cite{InW} and \cite{Co}).
Given the  decomposition (\ref{IW}), it was shown in \cite{InW}
that, the limit $C^k_{ij}(0)$ exists iff $e_i, i=1,\ldots,m$
span a subalgebra $W$ of $A$, which then characterizes the 
contraction. 

{Saletan} \cite{Sal} gives also a technical criterion for the 
existence of the limit 
$C^k_{ij}(0)$ defining his general contractions.
In contrast to the case of {In\"on\"u-Wigner} contractions,
a general {Saletan} contraction might be nontrivially iterated.

Not every transition $A\leq B$ 
corresponds to a  contraction.
While neither In\"on\"u-Wigner nor Saletan contractions 
are transitive, general transitions are transitive.
In this sense, the topological space $K^n$ provides the natural 
transitive completion of the well known contractions 
of In\"on\"u-Wigner and Saletan.
\section{\bf The classifying space $K^3$ of Bianchi Lie algebras}  
\setcounter{equation}{0}
The elements of $K^3$ are  well known
to correspond to the famous Bianchi Lie algebras,               
classified independently by Lie \cite{Lie} and Bianchi \cite{Bi}.  
For all types of Bianchi Lie algebras I up to IX an explicit description 
can be given in terms of the nonvanishing matrices  $C_{<i>}$,
$i=1,\ldots,3$, of some adjoint representation. 
This representation can be  
normalized modulo an overall scale of the basis $e_1, e_2, e_3$,
and moreover $C_{3}$ can be chosen in some normal form
(\cite{Rai1,Rai2} use the Jordan normal form).

In the {\em semisimple} representation category, there are only 
the simple Lie algebras $\VIII\equiv \so(1,2)=\su(1,1)$ and 
$\IX\equiv so(3)=su(2)$, given by 
\bear{SU2} 
&C_{<3>}(\VIII)=  
\left(
\begin{array}{cc}
 0  & 1 \\ 
 -1 & 0              
\end{array}
\right) ,
\qquad 
&C_{<1>}(\VIII) = -C_{<2>}(\VIII) = 
\left(
\begin{array}{cc}
 0     &  1 \\ 
  1 &  0              
\end{array}
\right) ,                                       
\nn\\
&C_{<3>}(\IX) = 
\left(
\begin{array}{cc}
 0  & 1 \\ 
 -1 & 0              
\end{array}
\right) ,
\qquad 
&C_{<1>}(\IX) = -C_{<2>}(\IX) = 
\left(
\begin{array}{cc}
 0     &  1 \\ 
 - 1 &  0              
\end{array}
\right) .                                       
\ear
All other algebras are in the {\em solvable} representation category.
They all have an Abelian ideal $\spn\{e_1,e_2\}$.
Hence, with vanishing $C_{<1>}=C_{<2>}=0$, 
they are described by $C_{<3>}$ only.
The "inhomogeneous" algebras $\VI_0\equiv \e(1,1)=\iso(1,1)$ 
(local isometry of a Minkowski plane) and 
$\VII_0=\e(2)=\iso(2)$ 
(local isometry of an Euclidean plane) are determined by  
\begin{equation}\label{E2} 
C_{<3>}(\VIo) = 
\left(
\begin{array}{cc}
 0  &  1 \\ 
 1  &  0              
\end{array}
\right) ,\qquad 
C_{<3>}(\VIIo) = 
\left(
\begin{array}{cc}
 0  &  1 \\ 
 -1  &  0              
\end{array}
\right) .                                       
\end{equation}
It holds $C_{<3>}(\VIII)=C_{<3>}(\VIIo)=C_{<3>}(\IX)$. 
Furthermore, $C_{<1>}(\VIII)=C_{<3>}(\VIo)$  and 
$C_{<1>}(\IX)=C_{<3>}(\VIIo)$.
So we find both transitions $\VIII\leq\VIIo$ and $\IX\leq\VIIo$,
but only  $\VIII\leq\VIo$.
These inhomogeneous algebras are endpoints $h=0$
of two $1$-parameter sets of algebras, $\VI_h$ and $\VII_h$,
for $h>0$ given respectively as   
\begin{equation} 
C_{<3>}(\VI_h) = 
\left(
\begin{array}{cc}
 h  &  1 \\ 
 1  &  h              
\end{array}
\right) ,\qquad 
C_{<3>}(\VII_h) = 
\left(
\begin{array}{cc}
 h   &  1 \\ 
 -1  &  h              
\end{array}
\right) .                                       
\end{equation}
Note also that  the $1$-parameter set of algebras $\VI_h$,
$0\leq h< \infty$ contains an
exceptional decomposable point $\III:=\VI_1=V^2\oplus \R$,
where $V^2$ is the unique non-Abelian algebra of $K^2$.
$C_{<3>}(\III)$ has exactly $1$ zero eigenvalue, while for all
other algebras $\VI_h$ and $\VII_h$, $0\leq h< \infty$ the
matrix $C_{<3>}$ has two different non-zero eigenvalues,
which become equal only in the limit $h\to\infty$.
If the geometric multiplicity of this limit is $1$, then
the latter corresponds to the Bianchi Lie algebra IV,
representable with
\beq{IV}
C_{<3>}(\IV) = 
\left(
\begin{array}{cc}
 1 & 1 \\ 
   & 1              
\end{array}
\right) .
\eeq

By {\em geometric specialization}   
of the algebra IV the geometric multiplicity of its
eigenvalue is increases to $2$, yielding the pure vector type algebra V,
given by
\beq{V}
C_{<3>}(\V) = 
\left(
\begin{array}{cc}
 1 &  \\ 
   & 1              
\end{array}
\right) .
\eeq

By {\em algebraic specialization}   
of the algebra IV the eigenvalue becomes zero, yielding the 
Heisenberg algebra II,
given by
\beq{II}
C_{<3>}(\II) = 
\left(
\begin{array}{cc}
 1 &  \\ 
   & 1              
\end{array}
\right) .
\eeq
This algebra can also be reached by via a direct transition from any of
the algebras $\VI_h$ and $\VII_h$, $0\leq h<\infty$. 
Finally both, geometric specialization of $\II$ and algebraic specialization
of $\V$, yield  the unique Abelian Lie algebra I, 
given by              
\beq{I}
C_{<i>}(\I) = 0, \qquad i=1,\ldots,3\ .
\eeq
\nl\nl\nl
\noindent
Fig. 1  shows on each horizontal level the algebras of equal index,
which are the sources for the level below, and possible targets
for the level above.
\vspace*{15.7truecm}
\nl\noindent
{\normalsize Fig. 1: The topological space $K^3$ (right and left images 
have to be identified for the algebras IV and V;
the locally maximal algebras IV, $\VI_h$ and $\VII_h$, $0\leq h<\infty$, 
form a $1$-parameter set of sources of arrows).}
\np
\nl\nl\nl
\noindent
Fig. 2 gives the analogous picture
for the space $K^3_{\C}$ of $3$-dimensional complex Lie algebras.
\vspace*{15.7truecm}
\nl\noindent
{\normalsize Fig. 2: The topological space $K^3_{\C}$ 
(the locally maximal algebras IV, $e_h$, $0\leq h<\infty$, form
a $1$-parameter set of sources of arrows).}
\np
\section{\bf Zariski dual topology and Lie algebra cohomology}  
\setcounter{equation}{0}
Now note that for any topology
there exists a {\em dual topology} by exchanging open and closed sets.
Applied to the topology $\kappa^n$, open points of $K^n$ become
closed and closed points become open for the dual topology.
Furthermore source and target of arrows interchange in the dual topology,
i.e. their arrows change their direction.
In the dual topology the rigid Lie algebras correspond to 
isolated closed points.
Actually all semisimple Lie algebras are such isolated closed points.

Recall now, that on any algebraic variety there is a unique topology, 
called the {\em Zariski topology}, such that its closed subsets correspond 
to algebraic subvarieties.
In this sense the topology $\kappa^n$ turns naturally out to be the  dual 
of the Zariski topology on $K^n$. So, what is the meaning of this
{\em Zariski dual topology} of $K^n$ 
and, more specifically of its {\em closed points} and 
{\em isolated closed points} ?
To answer that question, note first that the reversed arrows
of the dual topology correspond to some "inverse limit"
of the transitions along them. More generally the dual
of any transition might be called {\em spontaneous deformation} in $K^n$.
This has to be distinguished from a {\em parametrical deformation}
in $K^n$, which is given by a continuous  change of parameters within a 
Hausdorff connected component of $K^n$.
In the Zariski dual topology of $K^n$,
the closed points cannot be source of spontaneous deformations in $K^n$,
and the isolated closed points admit
neither spontaneous nor parametrical deformations.

Actually deformations of Lie algebras can also be considered
{}from a slightly different point of view, using
{\em Lie algebra cohomology}, introduced in \cite{CheE} and \cite{HochS}.
Let us consider a cochain complex 
\beq{complex}
{\ \atop 0} {\ \atop\to}{\ \atop C^0} {\delta_0 \atop\to} {\ \atop C^1} 
{\delta_1 \atop\to} {\ \atop C^2} {\ \atop\to}{\ \atop \ldots \ .}
\eeq
Its cochain spaces 
\beq{chain}
C^k(A,{\cal R}):=
\{ f: {\overbrace{A\otimes\cdots\otimes A}^{k}} \to {\cal R} \vert
f \ \mbox{linear, antisymmetric}  \}
\eeq
can generally be defined 
for any $A$-module $\cal R$ over some Lie algebra $A$.
The coboundary operators
are given by
\bear{coboundary}
\delta_k f(x_1,\ldots,x_n)
&:=&
\sum_{i=1}^{k} (-1)^{k+i} x_i f(x_1,\ldots,{\hat x}_i,\ldots,x_n)
\nn\\
& & +\sum_{i,j=1}^{k} (-1)^{i+j} 
f(x_1,\ldots,{\hat x}_i,\ldots,{\hat x}_j,\ldots,x_n,[x_i,x_j]) .
\ear
f is an {\em $k$-cocycle} iff $\delta_k f=0$. 
$Z^k(A,{\cal R}):=\ker \delta_k$.
f is an {\em $k$-coboundary} iff $f= \delta_{k-1} g$. 
$B^k(A,{\cal R}):=\im \delta_{k-1}$.
The coboundary (\ref{coboundary}) satisfies $\delta^2=0$,
hence $B^n(A,{\cal R})\subset Z^n(A,{\cal R})$ and 
the {\em $k^{th}$ cohomology} is defined
as $H^k(A,{\cal R}):=Z^k(A,{\cal R})/B^k(A,{\cal R})$.

Let us restrict now for simplicity to complexes (\ref{complex}) with
${\cal R}=A$, where the left multiplication by $x$ is just given 
by the adjoint action $\ad_x:=[x,\cdot]$, 
and write $C^k\equiv C^k(A,A)$ and $H^k\equiv H^k(A,A)$, keeping
in mind that these quantities all depend on the algebra $A\in K^n$.

The {\em deformation} $[\cdot,\cdot]_{\eps}$  of the 
product $[\cdot,\cdot]$ of some algebra $A$,
can be written as a formal power series
\beq{deform}
[x,y]_{\eps}=[x,y]+\eps F_1(x,y)+\eps^2 F_2(x,y)+\ldots .
\eeq
If the product  $[\cdot,\cdot]$ is defined in some category,
e.g. the category Lie products $W^n$ of Lie algebras in $K^n$,
the formally deformed product $[\cdot,\cdot]_{\eps}$ is in general
not well defined on the same category, but only on some extended
category. In order to be still a product in the same category,
here in $W^n$,
the deformed product $[\cdot,\cdot]_{\eps}$ 
has to satisfy an infinite number of 
{\em deformation equations},
namely for all $k\in \N_0$ the coefficients of the
formal power series have to satisfy
\beq{defeqn}
\sum_{x,y,z \atop \cyclic} \sum_{i+j=k} 
F_i(F_j(x,y),z)+F_j(F_i(x,y),z)=0 ,
\eeq
with $F_0\equiv [\cdot,\cdot]$.
For Lie products in $W^n$, the equation for $k=0$
corresponds to the Jacobi condition,
and the {\em infinitesimal deformation equation}, i.e. the equation
for $k=1$, can be expressed
with $\delta\equiv\delta_2$ from (\ref{coboundary}) as
\beq{infdef}
\delta F_1=0.
\eeq
So we see that $Z^2$ is the set of {\em infinitesimal deformations} 
of elements of $W^n$.
Some of these deformations 
yield the again the original algebra $A\in K^n$, 
i.e. they are deformations along
the $\GL(n)$-orbit through $F_0\in W^n$. 
These {\em trivial infinitesimal deformations} are elements
of $B^2$. Hence $H^2$ contains just the {\em non-trivial 
infinitesimal deformations}. If some algebra $A\in K^n$ satisfies
$H^2\equiv H^2(A,A)=0$, then this algebra cannot be source
of {\em infinitesimal deformations}. The latter may, corresponding
to the definitions above, be divided
into {\em infinitesimal spontaneous deformations} and
{\em infinitesimal parametrical deformations},
where the former are the duals of transitions and the latter
generate parametrical deformations within
the Hausdorff connected component of $A$. 
So, if  $A\in K^n$ is an isolated open point w.r.t. the original topology 
of $K^n$ or, equivalently, an isolated closed point w.r.t. the 
Zariski dual topology of $K^n$, then there are no non-trivial
infinitesimal deformations of the product $F_0\in W^n$; rather all
infinitesimal deformations are within $\pi^{-1}(A)\subset  W^n$. 
For these algebras $H^2=0$. In particular, the latter is known to 
be true for all semisimple Lie algebras.

\section{\bf Classifying local homogeneous Riemannian $3$-manifolds}  
\setcounter{equation}{0}
Now we will construct a classifying space for 
local homogeneous Riemannian $3$-manifolds.

The KS spaces appear as a limit 
of Bianchi IX spaces, in which the Bianchi IX isometry is still maintained, 
but no longer transitive. Hence it is sufficient to
consider local $3$-manifolds of Bianchi Lie isometry.

For the present case of Riemannian $3$-spaces $g_{ab}$ has a definite sign.
Let us consider the $3$-geometry {\em modulo} a transformation of the
its global sign,  
\beq{globalsign}
g_{ab}\to-g_{ab}.
\eeq
Then we can normalize the global sign with $\det(g_{ab})>0$
to $\epsilon_1=\epsilon_3=\epsilon_3=1$ in Eq. (\ref{gab}). 

The choice of real parameters $s, t, w$ of Eq. (\ref{gab})  
simplifies the following calculations in a specific 
triad basis for the Bianchi Lie algebras of Sec. 5.
This basis is chosen in consistency with the representations of 
\cite{LL} and \cite{Kr}.
It can be represented by
matrices $(e^a_\alpha)$, with anholonomic $a=1,2,3$ of the generators
of the algebra,
and holonomic coordinate columns $\alpha=1,2,3$.
The coordinates will also be denoted  as $x^1=:x$, $x^2=:y$, $x^3=:z$.
For the different Bianchi type these matrices take the following
form:
\nl
Bianchi I:
\begin{equation}
(e^a_\alpha)=
\left [
\begin {array}{ccc}        1&&
\\\noalign{\medskip}       &1&
\\\noalign{\medskip}       &&1  \end {array}
\right ],
\end{equation}
Bianchi II:
\begin{equation}
(e^a_\alpha)=
\left [
\begin {array}{ccc}        1&-z&
\\\noalign{\medskip}       0& 1&
\\\noalign{\medskip}       &&1  \end {array}
\right ],
\end{equation}
Bianchi IV:
\begin{equation}
(e^a_\alpha)=
\left [
\begin {array}{ccc}        1&&
\\\noalign{\medskip}       & e^x& 0
\\\noalign{\medskip}       &xe^x&e^x  \end {array}
\right ],
\end{equation}
Bianchi V:
\begin{equation}
(e^a_\alpha)=
\left [
\begin {array}{ccc}        1&&
\\\noalign{\medskip}       &e^x&
\\\noalign{\medskip}       &&e^x  \end {array}
\right ],
\end{equation}
Bianchi $\VI_h$, $h=A^2$:
\begin{equation}
(e^a_\alpha)=
\left [
\begin {array}{ccc}        1&&
\\\noalign{\medskip}       & e^{Ax}\cosh x  &-e^{Ax}\sinh x
\\\noalign{\medskip}       &-e^{Ax}\sinh x  & e^{Ax}\cosh x  \end {array}
\right ],
\end{equation}
Bianchi $\VII_h$, $h=A^2$:
\begin{equation}
(e^a_\alpha)=
\left [
\begin {array}{ccc}        1&&
\\\noalign{\medskip}       & e^{Ax}\cos x  &-e^{Ax}\sin x
\\\noalign{\medskip}       & e^{Ax}\sin x  & e^{Ax}\cos x  \end {array}
\right ],
\end{equation}
Bianchi VIII:
\begin{equation}
(e^a_\alpha)=
\left [
\begin {array}{ccc}        \cosh y \cos z &-\sin z& 0
\\\noalign{\medskip}       \cosh y \sin z & \cos z& 0 
\\\noalign{\medskip}       \sinh y        &    0  & 1  \end {array}
\right ],
\end{equation}
Bianchi IX:
\begin{equation}
(e^a_\alpha)=
\left [
\begin {array}{ccc}        \cos y \cos z &-\sin z& 0 
\\\noalign{\medskip}       \cos y \sin z & \cos z& 0  
\\\noalign{\medskip}       -\sin y       &    0  & 1  \end {array}
\right ].
\end{equation}
For each of the lines $\{\VI_h\}$ and $\{\VII_h\}$
the parameter range is given by ${\sqrt h}=A\in [0,\infty[$.
Let us also remind that, $\III:=\VII_1$.

The  structure constants 
can be reobtained from  $ds^2$ and the triad by 
\begin{equation}
C_{ijk}=ds^2([e_i,e_j],e_k),\qquad C^k_{ij}=C_{ijr}g^{rk}.
\end{equation}
The metrical connection  coefficients are determined as
\begin{equation}
\Gamma^k_{ij}=\frac{1}{2}  
g^{kr}(C_{ijr}+C_{jri}+C_{irj}).
\end{equation}
W.r.t. the triad basis, the curvature operator is defined as
\begin{equation}
\Re_{ij}:=\nabla_{[e_i,e_j]}
     -(\nabla_{e_i}\nabla_{e_j}-\nabla_{e_j}\nabla_{e_i}),
\end{equation} 
the Riemann tensor components are
\begin{equation}
R_{hijk}:=<e_h,\Re_{ij}e_k>,
\end{equation}
and the Ricci tensor is 
\beq{Ricci}                      
R_{ij}:=R^k_{ikj}
=\Gamma^f_{ij}\Gamma^e_{fe}-\Gamma^f_{ie}\Gamma^e_{fj}
	 +\Gamma^e_{if}C^f_{ej} .
\eeq
{}From (\ref{Ricci}) we may form the following scalar invariants
of the geometry:
The Ricci curvature scalar 
\begin{equation}
R:=R^i_{\ i}, 
\end{equation}
the sum of the squared eigenvalues 
\begin{equation}
N:=R^i_{\ j} R^j_{\ i},
\end{equation}
the trace-free scalar  
\begin{equation}
S:=S^i_{\ j} S^j_{\ k} S^k_{\ i}= R^i_{\ j} R^j_{\ k} R^k_{\ i} 
- R N + \frac{2}{9} R^3,
\end{equation} 
where $S^i_{\ j}:=R^i_{\ j}-\frac{1}{3}\delta^i_{\ j} R$, 
and, related to the York tensor,
\begin{equation}                  
Y:=R_{ik;j}g^{il}g^{jm}g^{kn}R_{lm;n}, \qquad \mbox{with}
\end{equation}
$$
R_{ij;k}:=e^\alpha_i e^\beta_j e^\gamma_k R_{\alpha\beta;\gamma} 
$$
$$
=e^l_{\alpha;\beta} e^\alpha_m e^\beta_k 
(\delta^m_i \delta^n_j + \delta^n_i \delta^m_j) R_{ln}. 
$$
The  $4$ scalar invariants above characterize a 
local homogeneous Riemannian $3$-space.

It is $N = 0$, iff the Riemannian $3$-space is the unique flat one.
This has a transitive isometry of Bianchi type I, and also 
admits the left-invariant (but not transitive) action  of the Bianchi  
group $\VII_0$ on its $2$-dimensional hyperplanes
(cf. \cite{Nom,Mil}).

In the following we take the flat Riemannian $3$-space as a center of 
projection for the non-flat Bianchi or KS geometries.
These satisfy $N\neq 0$. The invariant $N$ then parametrizes
(like $e^{-2s}$) the homogeneous conformal scale on the 
$3$-manifold under consideration. 
A homogeneous conformal, i.e. homothetic, rescaling  of 
the metric, 
\beq{rescale}
g_{ij} \to \sqrt{N} g_{ij},
\eeq                           
yields the following normalized invariants, which depend only on the
homogeneously conformal class of the geometry:
\begin{equation}\label{hat}                
{\hat N} := 1,
\quad
{\hat R} := R/{\sqrt{N}},
\quad
{\hat S} := S/N^{3/2},
\quad
{\hat Y} := Y/N^{3/2}.
\end{equation}
For a non-flat Riemannian space, the invariant ${\hat Y}$ vanishes, 
iff the $3$-geometry is conformally flat. Note that a general
conformal transformation is not necessarily homogeneous.
Hence there may exist homogeneous spaces, which are in the same
conformal class, but in different homogeneously conformal classes.

Note that, 
under (\ref{globalsign}),
$N$  is invariant,  
while $\hat R$, $\hat S$ and $\hat Y$ just all reverse their sign. 
Furthermore, a rescaling (\ref{rescale}) does not change the Bianchi
or KS type of isometry.

So we can now concentrate on the
classifying space of non-flat local homogeneous 
Riemannian $3$-geometries 
{\em modulo} the global sign (\ref{globalsign}) 
and 
{\em modulo} homogeneous conformal transformations (\ref{rescale}) 
for each fixed Bianchi type.
This {\em moduli space} can be parametrized by the invariants
$\hat R$, $\hat S$ and $\hat Y$, given for each fixed Bianchi type
as a function of the anisotropy parameters $t$ and $w$.   

A minimal cube, in which the classifying moduli space can be imbedded,
is spanned by
${\hat R}/\sqrt{3},\sqrt{6} {\hat S} \in [-1,1]$
and $2\tanh\hat Y\in[0,2]$.

Below, Fig. 3 describes those points of the moduli space
which are of Bianchi types VI/VII or lower level, 
Fig. 4 likewise points of Bianchi types VIII/IX. 
\np
\vspace*{19.7truecm}
\noindent
{\normalsize Fig. 3: Riemannian Bianchi geometries II, IV, V, $\VI_h(w=0)$,
\\ $\VI_h$
(${\sqrt h}=0,\frac{1}{5},\frac{1}{4},\frac{1}{3},\frac{5}{8},{1},{2}$), 
$\VII_h$
(${\sqrt h}=0,\frac{1}{7},\frac{1}{5},\frac{1}{4},\frac{1}{3},\frac{1}{2},{1}$);
\\
w.r.t. the common origin, the axes of the $3$ planar diagrams, are:
\\
$\hat R/\sqrt 3$ to the right, ${\sqrt 6}\hat S$ up, and $2\tanh \hat Y$ 
both, left and down.}
\np
\vspace*{19.7truecm}
\noindent
{\normalsize Fig. 4: Riemannian Bianchi geometries II, V, 
$\VI_0$, $\VI_1$, $\VII_0$, \\ 
$\VIII(t,w)$  ($t=-5,-1,0,1,5$),
$\IX(t,w)$
($t=0, \frac{1}{2}, 1, 2, 5$);
\\
w.r.t. the common origin, the axes of the $3$ planar diagrams, are:
\\
$\hat R/\sqrt 3$ to the right, ${\sqrt 6}\hat S$ up, and $2\tanh \hat Y$ 
both, left and down.}
\np
For a homogeneous space with $2$ equal {Ricci} eigenvalues 
the corresponding point in the ${\hat R}$-${\hat S}$-plane 
lies on a double line $L_2$,
which has a range defined by $\vert\hat R\vert \leq  \sqrt{3}$ 
and satisfies the algebraic equation
\begin{equation}\label{L2}
{162}{\hat S^2}=(3-\hat R^2)^3\ .            
\end{equation}  
All other algebraically possible points of the 
${\hat R}$-${\hat S}$-plane lie inside the region surrounded
by the line $L_2$.
At the branch points $\hat R  = \pm \sqrt 3$ of $L_2$ all Ricci eigenvalues 
are equal. 
These homogeneous spaces possess a $6$-dimensional isometry
group.
Homogeneous spaces possessing a $4$-dimensional isometry
group are represented by points on $L_2$.

If one Ricci eigenvalue equals $R$, i.e. 
if there exists a pair $(a,-a)$ of Ricci  eigenvalues,  
the corresponding point in the ${\hat R}$-${\hat S}$-plane 
lies on a line $L_{+-}$,
defined by the range $\vert\hat R\vert\leq 1$ 
and the algebraic equation
\begin{equation}\label{L+-}
{\hat S}=\frac{11}{9}\hat R^3-\hat R.            
\end{equation}  

In the case that one eigenvalue of the Ricci tensor is zero,
the corresponding point in the ${\hat R}$-${\hat S}$-plane lies 
on a line $L_{0}$, defined by the range $\vert\hat R\vert\leq\sqrt{2}$ 
and the algebraic equation
\begin{equation}\label{L0}
{\hat S}=\frac{\hat R}{2}(1-\frac{5}{9}\hat R^2).            
\end{equation}  

For Eqs. (\ref{L2}),(\ref{L+-}),(\ref{L0}) see also \cite{RaSch}.

At the branch points of the curve $L_2$ the Ricci  tensor has a 
triple eigenvalue, which is negative for geometries of Bianchi type V,
and positive for type $\IX$ geometries with parameters $(t,w)=(0,0)$.
These constant curvature geometries are all conformally flat with $\hat Y=0$.
Besides the flat Bianchi I geometry, the remaining conformally flat spaces 
with $\hat Y$ are the KS space  
$(\hat R,\hat S,\hat Y) = (\sqrt 2,- \frac{\sqrt 2}{18},0)$
and, point reflected,  the Bianchi type $\III_c$, corresponding to 
the initial point of a Bianchi III line segment ending at the Bianchi II
point in Fig. 3.

The point $(-1,0,0)$ of Fig. 3
admits both types, Bianchi V and $\VII_h$ with $h>0$. 
Nevertheless, this point corresponds only to one homogeneous  
space, namely the space of constant negative curvature.
This is possible, because this space has a $6$-dimensional
Lie group, which contains the  Bianchi V and $\VII_h$ subgroups.  
Note that in the flat limit $\V\to \I$, the additional Bianchi groups
$\VII_h$ change with $h\to 0$. 

Similarly, the Bianchi III points of Fig. 3 lie on the
curve $L_2$ of the $\hat R-\hat S$ diagram. 
However, these points are also of  
Bianchi type VIII. 
In fact, each of them correspond to one homogeneous geometry only. 
However, the latter admits a $4$-dimensional   
isometry group, which has two $3$-dimensional subgroups,
namely Bianchi III and VIII, both containing the same
$2$-dimensional non-Abelian subgroup.

Altogether, the location of Riemannian Bianchi (and KS)
spaces is consistent with the topology $\kappa^3$ of the space of
Bianchi Lie algebras.

Our classifying moduli space of local homogeneous
Riemannian $3$-spaces is a $T_2$ (Hausdorff) space. 
But it is not a topological manifold:
The line of $\VII_0$ moduli is a common boundary of
$3$ different $2$-faces, namely that of the IX moduli, that of the 
VIII moduli, and with $h\to 0$ that of all moduli of type $\VII_h$
with $h>0$.
Like the moduli space, also the full classifying space 
is not locally Euclidean; rather both are stratifiable varieties. 
\np

\section{\bf Discussion: Evolution of fundamental symmetries}  
\setcounter{equation}{0}
In this section we show up the possible application of our results 
to fundamental symmetries of physics. All applications typically involve
changes of the symmetry of the system under consideration; 
they differ in the kind of system that evolves the considered 
Lie symmetries. Here only two examples of fundamental systems
shall be mentioned:
The evolution of cosmological models and the connection dynamics of 
quantum gravity. 

The traditionally considered $1+3$-dimensional 
inhomogeneous cosmological models with homogenous Riemannian
$3$-hypersurfaces (see \cite{RaSch,Kras,Schm1}), 
and a more general class of multidimensional 
geometries, admit deformations between Riemannian Bianchi geometries.
Such a deformation may also induce a change of the 
spatial anisotropy of the universe, which essentially
affects physical quantities like its tunnelling rate
\cite{MM}.
Even for the more general multidimensional case, 
where $M=\R\times M_1\times\ldots\times M_n$ with homogeneous $M_1$, 
our results provide an important piece of information, 
namely the complete control
over the possible continuous deformations of the homogeneous external 
$3$-space $M_1$.

Furthermore, the superspace of homogeneous Riemannian $3$-geometries 
plays a key role for an understanding of the canonical quantization
of a homogeneous universe. The homogeneous conformal modes are just 
the homothetic scales, which span a $3$-dimensional minisuperspace
underlying the conformally equivariant quantization scheme
\cite{Rai3,Rai4}, yielding the 
Wheeler-deWitt equation for a given point of the moduli space 
of local homogeneous $3$-geometries. 
Global properties of the 
homogeneous $3$-geometry have not considered here. However it should 
be clear that a given global geometry according to one of the Thurston 
types exists only for a specific local geometries specified by 
characteristic points in our moduli space (cf. \cite{LRLu}).

The short distance regime of quantum gravity might be described
in terms of connection dynamics, recently also related to spin  
networks  \cite{RSm}. While the standard theory is worked out 
for the compact structure group $SU(2)$, the topology of $K^3$ suggests that 
this structure group could change by a transition to the 
noncompact group $E(2)$
and similarly further, until the $3$-dimensional Abelian group is reached. 
It remains an interesting question, what happens to connections, 
and moreover to the holonomy groups, under such a deformation.
In \cite{RSm} a q-deformation of $SU(2)$ was  suggested, in order to 
regularize infrared divergences. However, infrared divergences
are obviously related to the macroscopic limit. 
Since in this limit the dicrete structure of space-time
is expected to become replaced by a continuous structure,
we have no reason to expect that original spin network, or some
related braid structures, might be pertained in the macroscopic theory.
So a transition {\em within} the category of Lie algebras  
is more likely to provide the solution of the infrared problem, 
even more, since $K^3$-transitions appear naturally
in the cosmological evolution (e.g. related to the isotropization of
a homogeneous universe).

Finally it is suggestive that, the index technique
introduced in Sec. 3 might provide some kind of potential $J$ on $K^n$
determining the evolution of the symmetries under consideration.
For $n=3$, the $SU(2)$ symmetry would be {\em metastable}
and decay after some time to a state of  $K^3$ which has a lower potential
level, and so on,  until the minimal state of Abelian symmetry is reached.
The metastability of a higher level in the potential 
could have an explanation in the higher dimension of the corresponding  
subspaces of local homogeneous moduli of just that symmetry.
Note that for symmetries in $K^3$, 
the dimensionality of the subspace of corresponding moduli 
increases with the level of the potential.

\nl
\nl
\np \noindent
{\Large\bf Acknowledgement}
\nl
\nl
\setcounter{equation}{0}
The author thanks the DAAD for financial support, and the IPM for 
hospitality.
Discussions with  V. Karimipour, M. Mohazzab,
A. Mustafazadeh, H. Salehi and H.-J. Schmidt have contributed 
to clarify the present subject.  
\nl
\nl
{\Large\bf Appendix}
\nl
\nl
\setcounter{equation}{0}
\renewcommand{\theequation}{A.\arabic{equation}}
Here we list up the scalar invariants $\hat R$, $\hat S$ and $\hat Y$  
for all non-flat Riemannian Bianchi geometries:
\bear{BII}
\hat R_{\II}&=&
-{\frac {\sqrt {3}}{3}}
\nonumber\\
\hat S_{\II}&=&
{\frac {16\,\sqrt {3}}{81}}
\nonumber\\
\hat Y_{\II}&=&
{\frac {8\,\sqrt {3}}{9}}
\ear
\bear{BIV}
\hat R_{\IV}&=&
-{\frac {12\,{e^{w}}+1}{\sqrt {48\,{e^{2\,w}}+16\,{e^{w}}+3}}}
\nonumber\\
\hat S_{\IV}&=&
{\frac {16+72\,{e^{w}}}
{9\,\left (48\,{e^{2\,w}}+16\,{e^{w}}+3\right )^{3/2}}}
\nonumber\\
\hat Y_{\IV}&=&
{\frac {8+8\,{e^{w}}+32\,{e^{2\,w}}}
{\left (48\,{e^{2\,w}}+16\,{e^{w}}+3\right )
^{3/2}}}
\ear
\bear{BV}
\hat R_{\V}&=&
-\sqrt {3}
\nonumber\\
\hat S_{\V}&=&
0
\nonumber\\
\hat Y_{\V}&=&
0
\ear
In the next formulas an auxiliary invariant $D$ simplifies the notation.
\bear{VI}
D_{\VI_h}&:=& 3+4\,\left (4\,{h}+1\right ){e^{w}}+2\,\left (1+16\,{h}
+24\,{h}^{2}\right ){e^{2\,w}}
\nonumber\\ 
& & +4\,\left (4\,{h}+1\right ){e^{3\,w}}+3\,{e^{4\,w}} 
\nonumber\\
\hat R_{\VI_h}&=&-(D_{\VI_h})^{-\frac{1}{2}}
\left ( 1+2\,\left (1+6\,{h}\right ){e^{w}}+{e^{2\,w}} \right )
\nonumber\\
\hat S_{\VI_h}&=&\frac{8}{9}\, (D_{\VI_h})^{-\frac{3}{2}}
\left ({e^{w}}+1\right )^{4}\left (2+\left (9\,{h}-5\right )
{e^{w}}+2\,{e^{2\,w}}\right ) 
\nonumber\\
\hat Y_{\VI_h}&=& 8\, (D_{\VI_h})^{-\frac{3}{2}}
\left ({e^{w}}+1\right )^{2}\left ({e^{4\,w}}
+\left ({h}-1\right ){e^{3\,w}} \right.
\nonumber\\ 
& & \left.
+2\,\left ( 2\,{h}^{2} -5\,{h} +2 \right ) {e^{2\,w}}
+\left ( {h}-1  \right ) {e^{w}} + 1 \right )  
\ear
\bear{VII}
D_{\VII_h}&:=&3+4\,\left (4\,{h}-1\right ){e^{w}}
+2\,\left (1-16\,{h}+24\,{h}^{2}\right ){e^{2\,w}}
\nonumber\\ 
& &+4\,\left (4\,{h}-1\right ){e^{3\,w}}+3\,{e^{4\,w}}
\nonumber\\
\hat R_{\VII_h}&=& -(D_{\VII_h})^{-\frac{1}{2}}
\left ( 1+2\,\left (6\,{h}-1\right ){e^{w}}+{e^{2\,w}} \right)
\nonumber\\
\hat S_{\VII_h}&=& \frac{8}{9}\, (D_{\VII_h})^{-\frac{3}{2}} 
 \left ({e^{w}}-1\right )^{4}
\left (2+\left (9\,{h}+5\right ){e^{w}}+2\,{e^{2\,w}}\right )
\nonumber\\
\hat Y_{\VII_h}&=& 8\,  (D_{\VII_h})^{-\frac{3}{2}} 
\left ({e^{w}}-1\right )^{2}\left ({e^{4\,w}}
+\left ({h}+1\right ){e^{3\,w}} \right.
\nonumber\\
& & \left. 
+2\,\left (2\,{h}^{2}+5\,{h} +2\right ) {e^{2\,w}}
+\left ({h}+1\right ){e^{w}}+1\right )
\ear
\np\noindent
\bear{VIII}
D_{\VIII}&:=&
2\,{e^{-2\,w+2\,t}}+3\,{e^{-2\,w+4\,t}}+4\,{e^{-2\,w+t}}-4\,{e^{t}}
+4\,{e^{w}}+3\,{e^{2\,w}}
\nonumber\\
& & +4\,{e^{-w}}+3\,{e^{-2\,w}}-4\,{e^{-w+2\,t}}
+4\,{e^{-2\,w+3\,t}}+2+4\,{e^{-w+t}}
\nonumber\\
& & -4\,{e^{w+t}}
-4\,{e^{-w+3\,t}}+2\,{e^{2\,t}}
\nonumber\\
\hat R_{\VIII}&=& - (D_{\VIII})^{-\frac{1}{2}}
({2\,{e^{-w+t}}+{e^{-w}}+{e^{-w+2\,t}}+{e^{w}}+2-2\,{e^{t}}})
\nonumber\\
\hat S_{\VIII}&=& -\frac{8}{9}\, (D_{\VIII})^{-\frac{3}{2}} 
\left ( 6\,{e^{w+2\,t}}+3\,{e^{t+2\,w}}+3\,{e^{5\,t-2\,w}}
+14\,{e^{-3\,w+3\,t}} \right.
\nonumber\\
& & +6\,{e^{-3\,w+2\,t}}+6\,{e^{-3\,w+4\,t}}-3\,{e^{-3\,w+t}}
+6\,{e^{-w+4\,t}}-14\,{e^{3\,t}}
\nonumber\\
& & -3\,{e^{-3\,w+5\,t}}-2\,{e^{3\,w}}
-2\,{e^{-3\,w}}-15\,{e^{-w+t}}-3\,{e^{-2\,w}}
\nonumber\\
& & +15\,{e^{-2\,w+3\,t}}   +18\,{e^{w+t}}-15\,{e^{-w+3\,t}}-15\,{e^{2\,t}}
-3\,{e^{2\,w}}
\nonumber\\
& & +15\,{e^{t}}+18\,{e^{-2\,w+4\,t}}-18\,{e^{-2\,w+t}}+6\,{e^{-w}}
-42\,{e^{-w+2\,t}}
\nonumber\\
& & \left.
+6\,{e^{w}}-15\,{e^{-2\,w+2\,t}}-2\,{e^{-3\,w+6\,t}}+14 \right )
\nonumber\\
\hat Y_{\VIII}&=&  8\, (D_{\VIII})^{-\frac{3}{2}} 
\left ( 3\,{e^{w+2\,t}}-{e^{t+2\,w}}-{e^{5\,t-2\,w}}+6\,{e^{-3\,w+3\,t}}
 \right.
\nonumber\\
& & +3\,{e^{-3\,w+2\,t}}+3\,{e^{-3\,w+4\,t}}+{e^{-3\,w+t}}+3\,{e^{-w+4\,t}}
-6\,{e^{3\,t}}
\nonumber\\
& & +{e^{-3\,w+5\,t}}+{e^{3\,w}} +{e^{-3\,w}}-6\,{e^{-w+t}}
+{e^{-2\,w}}+6\,{e^{-2\,w+3\,t}}
\nonumber\\
& &
+5\,{e^{w+t}}-6\,{e^{-w+3\,t}}-6\,{e^{2\,t}}
+{e^{2\,w}}+6\,{e^{t}}+5\,{e^{-2\,w+4\,t}}
-5\,{e^{-2\,w+t}}
\nonumber\\
& & \left.
+3\,{e^{-w}}-18\,{e^{-w+2\,t}}+3\,{e^{w}}
-6\,{e^{-2\,w+2\,t}}+{e^{-3\,w+6\,t}}+6 \right )
\ear
\bear{IX}
D_{\IX}&:=&
2+3\,{e^{2\,w}}+3\,{e^{-2\,w}}-4\,{e^{-w}}-4\,{e^{-2\,w+t}}
+3\,{e^{-2\,w+4\,t}}+2\,{e^{2\,t}}
\nonumber\\
&&+2\,{e^{-2\,w+2\,t}}+4\,{e^{-w+2\,t}}
-4\,{e^{-2\,w+3\,t}}-4\,{e^{w}}+4\,{e^{t}}
\nonumber\\
&&+4\,{e^{-w+t}}-4\,{e^{-w+3\,t}}
-4\,{e^{w+t}}
\nonumber\\
\hat R_{\IX}&=&  (D_{\IX})^{-\frac{1}{2}}
\left ( -{e^{w}}-{e^{-w+2\,t}}-{e^{-w}}+2\,{e^{-w+t}}+2+2\,{e^{t}}
\right )
\nonumber\\
\hat S_{\IX}&=& -\frac{8}{9}\,  (D_{\IX})^{-\frac{3}{2}} 
\left( 6\,{e^{w+2\,t}}+3\,{e^{t+2\,w}}+3\,{e^{5\,t-2\,w}}
-14\,{e^{-3\,w+3\,t}} \right.
\nonumber\\
&&+6\,{e^{-3\,w+2\,t}}+6\,{e^{-3\,w+4\,t}}
+3\,{e^{-3\,w+t}}+6\,{e^{-w+4\,t}}-14\,{e^{3\,t}}
\nonumber\\
&&+3\,{e^{-3\,w+5\,t}}
-2\,{e^{3\,w}}-2\,{e^{-3\,w}}+15\,{e^{-w+t}}
+3\,{e^{-2\,w}}
\nonumber\\
&&+15\,{e^{-2\,w+3\,t}}-18\,{e^{w+t}}+15\,{e^{-w+3\,t}}
+15\,{e^{2\,t}}+3\,{e^{2\,w}}
\nonumber\\
&&+15\,{e^{t}}-18\,{e^{-2\,w+4\,t}}
-18\,{e^{-2\,w+t}}+6\,{e^{-w}}-42\,{e^{-w+2\,t}}
\nonumber\\
&& \left.
+6\,{e^{w}}
+15\,{e^{-2\,w+2\,t}}-2\,{e^{-3\,w+6\,t}}-14 \right )
\nonumber\\
\hat Y_{\IX}&=& -8\, (D_{\IX})^{-\frac{3}{2}} 
\left( -3\,{e^{w+2\,t}}+{e^{t+2\,w}}+{e^{5\,t-2\,w}}+6\,{e^{-3\,w+3\,t}}
\right.
\nonumber\\
&& -3\,{e^{-3\,w+2\,t}}-3\,{e^{-3\,w+4\,t}}+{e^{-3\,w+t}}-3\,{e^{-w+4\,t}}
+6\,{e^{3\,t}}
\nonumber\\
&&+{e^{-3\,w+5\,t}}-{e^{3\,w}}-{e^{-3\,w}}-6\,{e^{-w+t}}
+{e^{-2\,w}}-6\,{e^{-2\,w+3\,t}}
\nonumber\\
&&+5\,{e^{w+t}}-6\,{e^{-w+3\,t}}-6\,{e^{2\,t}}+{e^{2\,w}}-6\,{e^{t}}
+5\,{e^{-2\,w+4\,t}}+5\,{e^{-2\,w+t}}
\nonumber\\
&&\left. 
-3\,{e^{-w}}
+18\,{e^{-w+2\,t}}-3\,{e^{w}}-6\,{e^{-2\,w+2\,t}}
-{e^{-3\,w+6\,t}}+6 \right)
\ear
\np\noindent

\end{document}